\documentclass[onecolumn,amsmath,amssymb,amsfonts,a4paper,aps,prd,10pt]{revtex4}
\usepackage{graphicx}
\usepackage{epstopdf}
\usepackage{verbatim}
\newcommand{\nc}{\newcommand}
\nc{\ba}{\begin{eqnarray}} \nc{\ea}{\end{eqnarray}}
\newcommand\be{\begin{equation}}
\newcommand\ee{\end{equation}}
\nc{\D}{\overline{\mbox{D3}}}

\nc{\ga}{\gamma} \nc{\tnu}{\tilde{\nu}} \nc{\tmu}{\tilde{\mu}}

\nc{\x}{{\bf{x}}}

\input{epsf.sty}
\begin{document}

%%%%%%%%%%%%%%%%%%%%%%%%%%%%%%%%%%%%%%%%%%%%%%%%%%%%%%%%%%%%%%%%%%%%%%%
%\title{Multi-Connection Framework\\an example: massive gravity from Weyl geometry}
%\title{Multi-Connection Framework\\a possible geometrical meaning for massive gravity}
%\title{Massive Gravity in Multi-Connection Formalism:\\Geometrical Realization of Graviton's Mass}
%\title{Multi-Connection Formalism:\\Inherent Tension of Space-Time a Realization of Massive Gravity}
\title{Geometric Massive Gravity in Multi-Connection Framework}

\author{Nima Khosravi}
\email{nima@aims.ac.za}

\affiliation{Cosmology Group, African Institute for Mathematical
Sciences, Muizenberg 7945, Cape Town, South Africa}

\begin{abstract}

\begin{comment}

We introduce a new framework for general relativity by allowing
existence of more than one connection on the manifold. By a
redefinition of variables the given connections can be written as
their average and their differences which are connection and tensors
respectively. At the kinematic level, the average connection plays the
role of connection in the geodesic equation. The simplest choice for
the dynamics is employed which is Einstein-Hilbert action. To do
this we need an auxiliary field (metric) to make the Lagrangian
scalar density. The equation of motion imposes that the average
connection be the Christoffel symbol corresponding to the metric.
However the difference tensors can be studied if they depend on at
least a part of the metric. In contrast to multi-metric models,
this framework has by construction, no ambiguities in definition
of the physical metric and consequently in the coupling to matter. As an
example we study a model inspired by Weyl geometry on a
bi-connection framework. In this example the metric compatibility
relation is broken perturbatively. It is shown that the Fierz-Pauli
mass term can be deduced automatically without any a priori tuning.
It is possible to generalize this result to de Rham-Gabadadze-Tolley
model.

\end{comment}

What is the right way to interpret a massive graviton? We generalize the kinematical framework of general relativity to multiple connections. The average of the connections is itself a connection and plays the role of the canonical connection in standard General Relativity. At the level of dynamics, the simplest choice of the Einstein-Hilbert action is indistinguishable from the single-connection case. However, inspired by Weyl geometry, we show how one can construct massive gravity to all orders in perturbation theory compatible with the de Rham-Gabadadze-Tolley ghost-free model. We conclude that the mass of the graviton can be interpreted as a geometrical property of spacetime arising from two connections. Furthermore in the multi-connection framework there is no ambiguity in the definition of physical metric and
consequently  coupling to matter.

\end{abstract}
\maketitle
%\tableofcontents
%%%%%%%%%%%%%%%%%%%%%%%%%%%%%%%%%%%%%%%%%%%%%%%%%%%%%%%%%%%%%%%%%%%%%%%%%%%%%%%%%%%%%%%%%%%%
%\newpage
\section{Introduction and Motivations}
\subsection{Generalization of Kinematics}
Einstein general relativity (EGR) has had a lot of successes
theoretically and observationally. Einstein (and Hilbert) suggested
a dynamical theory for the metric of spacetime. This theory improved
our knowledge not only of gravitational force dynamics but also in
its geometrical interpretation. At the kinematical level the EGR is
based on a manifold with a  metric living on it. This metric is
responsible for all the geometrical properties of spacetime. It
manifests the notion of distance, causal structure and parallel
transportation. The latter is realized by the corresponding connection
to the given metric i.e. Christoffel symbol. However it should be
emphasized that in a general framework the connection can be an
independent geometrical object which is responsible for parallel
transportation and consequently the geodesic equation and covariant
derivative. The Christoffel symbol is a specific kind of connection
which satisfies the metric compatibility relation.  It is a unique property of
Einstein-Hilbert action that imposes a priori independent connection have to be the
Christoffel symbol according to the equations of motion.

Though  EGR can be interpreted as a triumph in our understanding of
gravitational force, there have always been attempts at 
modifications \cite{pedro}. One reason for that is the problem of
explaining observational data such as dark energy or
dark matter\footnote{About dark matter there is a very strong
alternative from particle physics viewpoint.}. Other reasons include
 theoretical challenges e.g. cosmological constant problem, singularities and quantum gravity, as
well as curiosity of theoretical physicists. For this purpose, one
method is a modification at the level of dynamics i.e. the
Einstein-Hilbert action e.g. $f(R)$ or $R_{\mu\nu}R^{\mu\nu}$. Such
models usually share a kinematic structure with EGR i.e.
existence of a metric field and its corresponding Christoffel
symbol. However it has been shown that in general, dynamics does not
impose metric compatibility. An alternative approach\footnote{It should be
emphasized that both approaches result in the same physical consequence which
is the additional degrees of freedom. For example see N. Arkani-Hamed's talk
in Prospects in Theoretical Physics (PiTP) 2011.} to modify EGR
can happen at the level of kinematics e.g. various bi-metric models
\cite{bi-metric}. Bi-metric models are interesting to study because
their richer foundation allows for more opportunities at least
theoretically e.g. massive gravity \cite{bimetric-massive}. But
existence of more than one metric on the manifold causes some
ambiguities. For example our understanding of the measurement of
distance, the causal structure of manifolds and the geodesic equation are not
uniquely defined in this model. In addition there is another
challenge in coupling of matter and gravity which is crucial to
understand the behavior of the model in real circumstances.

In this work we introduce a new gravity model with modified kinematics. We do this by
allowing the manifold to have more than one connection. We will show
how this model has the advantages of its predecessor (i.e. a
multi-metric model) but without ambiguity in its physical
interpretation. As an example we will study massive gravity in this
framework, and so next present a brief review on the status of
massive gravity.

\subsection{Massive Gravity}
Recently massive gravity has attracted a lot of attention due to a milestone in this topic by de Rham,
Gabadadze and Tolley (dRGT) \cite{massivegravity}. They could
improve existence of a ghost free massive gravity which is an
extension of the Fierz-Pauli \cite{oldmassivegravity} massive
gravity to a non-linear regime. In dRGT massive gravity the
Boulware-Deser ghost \cite{b-deser} is absent. The dRGT massive
gravity needs a fiducial metric in addition to physical metric to
construct the mass (potential) term. This fact is easy to
understand since the potential in EGR should be a scalar made of
just the metric without any derivative operator. The only
possibility is the cosmological constant term. So to go further e.g.
a mass term, having two metrics seems an essential assumption. In
this sense massive gravity can be categorized as a bi-metric model.
In addition to  the problems of bi-metric model there is a fundamentally important question in massive gravity:
What is the \textit{geometrical} meaning of the mass of graviton? We
will try to address this question in the multi-connection framework.
But before that let us illustrate why we think this question is
important.

After Einstein, not only our knowledge about gravitational force
became improved but also it changed our viewpoint on the
interpretation of the gravitational force. EGR says very briefly
that gravity is geometry which is very profound. EGR formalism
behaves with gravity as a field theory same as e.g. electrodynamics.
It has been shown that field theoretical viewpoint on EGR results in
a massless spin-two particle named graviton which is responsible for
gravity force exchange same as photon in electrodynamics. Then the
lack of accurate observational data makes it possible to ask if the
graviton have a mass? While this can be a quite straightforward question
in particle physics, its realization theoretically and its
observational consequences should be carefully considered. As we
mentioned above the dRGT model has proposed a
well-defined massive gravity without any ghosts. In this formalism, as
mentioned in \cite{bimetric-massive}, all the realizations are
equivalent to assuming a bi-metric model in a four dimensional
geometry. It is worth mentioning that the bi-metric models have had
their own history \cite{bi-metric} though became more interesting
after establishing their relation with massive gravity. So one can
expect that the problems of bi-metric models exist in massive
gravity too. In this work we are going to study massive gravity as
an example in a multi-connection framework.

\section{Multi-Connection framework}
The  framework has a lot in common with multi-metric
models, but it will be shown that multi-connection framework is more
straightforward for physical interpretations. In addition it seems
this framework does not suffer from the usual problems of
multi-metric models mentioned earlier. To start we study the
kinematics of this framework and then consider its dynamics.

\subsection{Kinematics}
At the level of kinematics, a connection is responsible for parallel
transportation and consequently shows itself in the geodesic equation\footnote{Note that
we do not assume geodesic equation as a result of variation of an action i.e. $S=\int \sqrt{g_{\mu\nu}dx^\mu dx^\nu}$. This assumtion says the connection is the Christofell symbol.}.
The geodesic equation can be written as
\begin{eqnarray}\label{geodesic-eqn-rewritten}
\frac{d^2x^\mu}{d\lambda^2}=-\Gamma^{\mu}_{\alpha\beta}\frac{dx^\alpha}{d\lambda}\frac{dx^\beta}{d\lambda}.
\end{eqnarray}
In the above equation the left hand side is simply the
acceleration which appears in the Newton's second law as
$a^\mu=\frac{d^2x^\mu}{d\lambda^2}=\frac{1}{m}F^\mu$ where $\lambda$
is time, $m$ is mass and $F^\mu$ is an external force. By this
viewpoint the above equation can be interpreted as the Newton's
second law by assuming $F_{geo.}^\mu\propto
-\Gamma^{\mu}_{\alpha\beta}\frac{dx^\alpha}{d\lambda}\frac{dx^\beta}{d\lambda}$
and interpreting $F_{geo.}^\mu$ as a geometrical force. Now it is
straightforward to add new connections by reminding what happens to
Newton's second law when we have more than one forces. The geodesic
equation becomes
\begin{eqnarray}\label{geodesic-eqn-generalized}
\frac{d^2x^\mu}{d\lambda^2}=\sum_{i=1}^N{^{(i)}}F_{geo.}^\mu=-\sum_{i=1}^N
{^{(i)}}\Gamma^{\mu}_{\alpha\beta}\frac{dx^\alpha}{d\lambda}\frac{dx^\beta}{d\lambda}=
-N\left(\frac{1}{N}\sum_{i=1}^N
{^{(i)}}\Gamma^{\mu}_{\alpha\beta}\right)\frac{dx^\alpha}{d\lambda}\frac{dx^\beta}{d\lambda}
=-N\gamma^{\mu}_{\alpha\beta}\frac{dx^\alpha}{d\lambda}\frac{dx^\beta}{d\lambda},
\end{eqnarray}
where  each connection is labeled by $(i)$ and we define the
average connection as $\gamma^{\mu}_{\alpha\beta}\equiv
\frac{1}{N}\sum_{i=1}^N {^{(i)}}\Gamma^{\mu}_{\alpha\beta}$. It is
worth remembering that the average of connections is a connection
itself. So effectively the geodesic equation\footnote{Note that the
extra $N$ in the last term can be absorbed  by rescaling the coordinates as
$x^\mu\rightarrow\frac{1}{N}x^\mu$.}
(\ref{geodesic-eqn-generalized}) in the presence of many connections
is the geodesic equation in (\ref{geodesic-eqn-rewritten}) where the
average connection, $\gamma^{\mu}_{\alpha\beta}$, plays the role of
the canonical connection. As a consequence the covariant derivative should be
defined due to average connection $\gamma^{\mu}_{\alpha\beta}$ to be
compatible with the definition of the geodesic equation. So the
multi-connection framework works properly at the kinematic level.
Now let us turn to dynamics of the model where  the average
connection will show itself again.

\subsection{Dynamics}
Now having reviewed the kinematics of this model, we are next going to study
its dynamics. To construct the Lagrangian for this model we need an
auxiliary field to make scalars. By having this field $g_{\mu\nu}$ \footnote{We will see this
auxiliary field will play the role of the metric.} and consequently
its inverse $g^{\mu\nu}$ and determinant $g$ we can write the
simplest action inspired by Einstein-Hilbert action as
\begin{eqnarray}\label{lag-multi-conn}
{\cal{S}}=\int d^4x \sqrt{-g}g^{\mu\nu}\frac{1}{N}\sum_{i=1}^N
R_{\mu\nu}\left(^{(i)}\Gamma^\rho_{\alpha\beta}\right).
\end{eqnarray}
For our purpose in this work we restrict the model to a
bi-connection model. This assumption is just for simplicity and does
not change the physical consequences. The above Lagrangian
reduces to
\begin{eqnarray}\label{lag-two-conn}
{\cal{S}}=\int d^4x {\cal{L}}=\int d^4x
\sqrt{g}g^{\mu\nu}\frac{1}{2}\left[
R_{\mu\nu}\left(^{(1)}\Gamma^\rho_{\alpha\beta}\right)+
R_{\mu\nu}\left(^{(2)}\Gamma^\rho_{\alpha\beta}\right)\right]
\end{eqnarray}
which can be written as follows 
\begin{eqnarray}\label{lag-two-conn-tensor}
{\cal{L}}=\sqrt{g} g^{\mu\nu}
\left[R_{\mu\nu}\left(\gamma^\rho_{\alpha\beta}\right)+\Omega^{\alpha}_{\alpha\lambda}
\Omega^{\lambda}_{\nu\mu}-\Omega^{\alpha}_{\nu\lambda}\Omega^{\lambda}_{\alpha\mu}\right]
\end{eqnarray}
where $R_{\mu\nu}(\gamma^\rho_{\alpha\beta})$ is the Ricci tensor defined by
 the average connection $\gamma^\rho_{\alpha\beta}\equiv\frac{1}{2}\left(^{(1)}\Gamma^\rho_{\alpha\beta}+
^{(2)}\Gamma^\rho_{\alpha\beta}\right)$   and
$\Omega^\rho_{\alpha\beta}\equiv\frac{1}{2}\left(^{(1)}\Gamma^\rho_{\alpha\beta}-
^{(2)}\Gamma^\rho_{\alpha\beta}\right)$ is a tensor due to
transformation rule of connections. Since we assume connections are
symmetric in their lower indexes so $\gamma^\rho_{\alpha\beta}$ and
$\Omega^\rho_{\alpha\beta}$ are both symmetric in their lower
indexes. The variation of the Lagrangian with respect to
$\gamma^\rho_{\alpha\beta}$ 
results in
\begin{eqnarray}\label{metric-connection-2}
\gamma^{\rho}_{\alpha\beta}=\frac{1}{2}g^{\rho\mu}\left(\partial_\alpha
g_{\mu\beta}+\partial_\beta g_{\mu\alpha}-
\partial_\mu g_{\alpha\beta}\right),
\end{eqnarray}
which means $\gamma^{\rho}_{\alpha\beta}$ is a metric compatible
connection. So effectively the above Lagrangian can be written as
\begin{eqnarray}\label{lag-metric-tensor}
{\cal{L}}=\sqrt{g} g^{\mu\nu}
\left[R_{\mu\nu}\left(g_{\alpha\beta}\right)+\Omega^{\alpha}_{\alpha\lambda}
\Omega^{\lambda}_{\nu\mu}-\Omega^{\alpha}_{\nu\lambda}\Omega^{\lambda}_{\alpha\mu}\right]
\end{eqnarray}
where $g_{\mu\nu}$ and $\Omega^{\alpha}_{\mu\nu}$ can be independent
fields\footnote{It is worth to mention that in the Lagrangian (\ref{lag-metric-tensor}) the last term i.e. $\Omega^{\alpha}_{\alpha\lambda}
\Omega^{\lambda}_{\nu\mu}-\Omega^{\alpha}_{\nu\lambda}\Omega^{\lambda}_{\alpha\mu}$ is exactly $\Upsilon_{\mu\nu}$ in bi-metric MOND model \cite{mond}. However in \cite{mond} this term has been chosen by hand but here this structure appears automatically. We emphasize that bi-metric MOND model is a bi-metric model and fundamentally it is different with our model.}. However we should be very careful about this assumption. If
$g_{\mu\nu}$ and $\Omega^{\alpha}_{\mu\nu}$ are totally independent
then the equation of motion with respect to
$\Omega^{\alpha}_{\mu\nu}$ imposes $\Omega^{\alpha}_{\mu\nu}=0$.
This result can be seen more directly from the Lagrangian
(\ref{lag-two-conn}) by applying Palatini method on both
$^{(1)}\Gamma^\rho_{\alpha\beta}$ and
$^{(2)}\Gamma^\rho_{\alpha\beta}$. Then the result says
$^{(1)}\Gamma^\rho_{\alpha\beta}=^{(2)}\Gamma^\rho_{\alpha\beta}=\gamma^\rho_{\alpha\beta}$
where $\gamma^\rho_{\alpha\beta}$ is the Christoffel symbol
(\ref{metric-connection-2}). Therefore assuming totally independent
$g_{\mu\nu}$ and $\Omega^{\alpha}_{\mu\nu}$ reduces the model to
pure EGR.

However it is possible to do further analysis by a delicate
assumption and a generalization of the Einstein-Hilbert inspired action.
They are many possible candidates which all are interesting to study
and we mention some of them. I) One can modify the action
(\ref{lag-two-conn}) by assuming $f(R)$ inspired models. In this
case the Palatini method does not impose all the connections to be
Christoffel symbols and allows the non-vanishing
$\Omega^{\alpha}_{\mu\nu}$. II) It is possible to add a kinetic term
for $\Omega^{\alpha}_{\mu\nu}$ by hand and consider the
model. In
principle it is same as EGR in presence of another field e.g. in
Brans-Dicke model. III) The other candidate is breaking somehow the
independence between $g_{\mu\nu}$ and $\Omega^{\alpha}_{\mu\nu}$. As
it is obvious for each candidate there are many kinds of realizations
but in this work we consider the last choice and we will see how we
can break the mentioned independence employing  Weyl geometry.

\subsection{Coupling to Matter}
It is always a problem of bi-metric models that what is the role of
each metric i.e. which one (or combination) is the physical metric. One way to ask this
question is to ask which combination of the given metrics is
responsible for coupling to matter. Note that the answer to this
question can address the other related problems such as definition
of distance and causality in bi-metric models. In the
multi-connection framework this problem should exist but there is a
unique natural answer to it. This unique candidate is based on
assuming indistinguishability of the priory given connections and is very
straightforward to see because of the model's construction. By
looking at the Lagrangian (\ref{lag-two-conn}) it is natural to
think that the matter is coupled to $g_{\mu\nu}$ the auxiliary field
which is the metric. It is
very important to emphasize that this claim is compatible with how the
multi-connection framework behaves at the kinematic level. By
looking at the geodesic equation in this framework
(\ref{geodesic-eqn-generalized}) and having in mind the equivalence
principle it is easy to say that the matter should see the average
connection $\gamma^\alpha_{\mu\nu}$. It is exactly equivalent to say
matter is coupled to the metric $g_{\mu\nu}$ because we have shown
that the average connection $\gamma^\alpha_{\mu\nu}$ is the
Christoffel symbol according to the metric $g_{\mu\nu}$. According
to above arguments there is just one metric in our model which is
responsible for coupling to matter and so it is in charge for
measuring the distance and the causal structure too. It means
multi-connection framework has no problem with these
issues.

From this natural and unique result in multi-connection framework we
can have a hint about multi-metric gravity models. Suppose we start
with a bi-metric model then for each metric we can associate a
connection (Christoffel symbol) and the average connection can be
defined consequently. According to our result this average
connection is responsible for the coupling to matter in the geodesic
equation. However it is not trivial if is always possible to associate a metric to this average connection.

\begin{comment}
So if one can associate a metric to the average connection
then that metric is in charge to couple to matter in the Lagrangian.
However it is not trivial if it is always possible to correspond a
metric to a connection which is the average of two other
connections. These latter connections are Christoffel symbols with
respect to the two given metric in a bi-metric model.
\end{comment}

%\begin{comment}

\subsection{An Example}
Now let's assume a special kind of definition for
$\Omega^\alpha_{\mu\nu}$ as an example to bridge between
multi-connection formalism and massive gravity. This special form is
as follows
\begin{eqnarray}\label{massive-omega}
\Omega_{\alpha\mu\nu}\equiv \frac{1}{2}(C_\mu X_{\nu\alpha}+
C_\nu X_{\mu\alpha}-C_\alpha X_{\mu\nu})
\end{eqnarray}
where $X_{\mu\nu}$ is a symmetric tensor and $C_{\mu}$ is a
vector. With the above form for $\Omega^\alpha_{\mu\nu}$ the second
term of the Lagrangian (\ref{lag-metric-tensor}) reduces to
\begin{eqnarray}\label{lag-massive-1}
{\cal{L}}_{MG}=\frac{1}{4}\times\left[ 2(X
X_{\mu\nu}-X_{\mu\alpha}X^{\alpha}_\nu)C^\mu
C^\nu+(X_{\mu\nu}X^{\mu\nu}-X^2)C^2\right]
\end{eqnarray}
where  $X=g^{\mu\nu}X_{\mu\nu}$, $C^2=C_\mu C^\mu$ and
$g_{\mu\nu}$ is responsible for lowering and raising the indices.
For a special case which $C^\mu$ satisfies $C^2=-m^2$ and $C^\mu
X_{\mu\nu}=0$ the above Lagrangian becomes\footnote{In a
special case where $\eta_{\mu\nu}$ is the background metric, one can
assume a Gaussian distribution for $C_\mu$ then can use $\langle
C_\mu C_\nu\rangle=C^2\eta_{\mu\nu}$. This fact transforms the
Lagrangian (\ref{lag-massive-1}) to (\ref{lag-massive-2}) by
assuming $C^2=m^2$.}
\begin{eqnarray}\label{lag-massive-2}
{\cal{L}}_{MG}=-\frac{m^2}{4}\left(X_{\mu\nu}X^{\mu\nu}-X^2\right).
\end{eqnarray}
The important point about this Lagrangian is the relative
coefficient between $X_{\mu\nu}X^{\mu\nu}$ and
$X^2$. We emphasize that this relative coefficient is not
trivial and is related to the absence of ghosts in massive
gravity. We did not say what is the reason for taking the above
specific form for $\Omega^\alpha_{\mu\nu}$ in (\ref{massive-omega}).
In the next section we show how Weyl geometry can inspire us to find
a physical meaning for the above form of $\Omega^\alpha_{\mu\nu}$
for a specific $X_{\mu\nu}$.

But before that it is worth to mentioning that the above form of
the Lagrangian reminds us the dRGT massive gravity. By looking at
$X_{\mu\nu}\equiv{\cal{K}}_{\mu\nu}$ as the tensor defined in dRGT massive gravity 
\cite{massivegravity} i.e.
${\cal{K}}^\mu_\nu=\delta^\mu_\nu-\sqrt{\delta^\mu_\nu-H^\mu_\nu}$
where one can define $H_{\mu\nu}$ by using
$g_{\mu\nu}=\eta_{\mu\nu}+h_{\mu\nu}=H_{\mu\nu}+\eta_{ab}\partial^a
\phi_\mu \partial^b\phi_\nu$ where $\phi_\mu$ are Stukelberg fields.  So it can be concluded
that the diﬀerence between diﬀerent connections can show itself as a mass term for the graviton.  However to make a fully comparison with dRGT model we need to consider higher order terms ${\cal O}({\cal K}^3)$ and ${\cal O}({\cal K}^4)$. We will come back to this issue in the next section.

%\end{comment}

\section{Weyl Geometrical Massive Gravity}
In this section we will show how by employing Weyl geometry (WG) in
the multi-connection framework we can find a geometrical realization
of massive gravity. WG is an important generalization of Riemannian
geometry. In a classic work by Ehlers, Pirani and Schild it has been
claimed that WG can be deduced from an axiomatical approach to
general relativity \cite{pirani}. In WG in addition to the metric there is
another geometrical object which is a vector. This vector changes
the amplitude of a given vector due to parallel transportation
which is an additional effect to changing in the direction which happens
in Riemannian geometry. In WG the connection can be written as
\begin{eqnarray}\label{connection-WG}
\Gamma^\alpha_{\mu\nu}=\big\{^\alpha_{\mu\nu}\big\}-\frac{1}{2}g^{\alpha\beta}(
g_{\nu\beta}C_\mu+  g_{\mu\beta}C_\nu- g_{\mu\nu}C_\beta)
\end{eqnarray}
where $\Gamma^\alpha_{\mu\nu}$ is the connection and
$\big\{^\alpha_{\mu\nu}\big\}$ is the Christoffel symbol. This means
the metric compatibility relation modifies to $\nabla_\alpha
g_{\mu\nu}=C_\alpha g_{\mu\nu}$ where the covariant derivative is
due to $\Gamma^\alpha_{\mu\nu}$ and which obviously reduces to
Riemannian geometry for $C_\alpha=0$. However it is easy to show
that
\begin{eqnarray}\label{generalized-connection-WG}
\Gamma^\alpha_{\mu\nu}=\big\{^\alpha_{\mu\nu}\big\}-\frac{1}{2}g^{\alpha\beta}(
A_{\nu\beta}C_\mu+  A_{\mu\beta}C_\nu-
A_{\mu\nu}C_\beta)\hspace{1cm}\Longrightarrow\hspace{1cm}\nabla_\mu
g_{\alpha\beta}=C_\mu A_{\alpha\beta}
\end{eqnarray}
for a symmetric arbitrary tensor $A_{\alpha\beta}$. Inspired by
above arguments we are going to utilize the above results in the
multi-connection framework. Now assume a special case in our
bi-connection model as
follows
\begin{eqnarray}\label{2-connection-WG}
^{(1)}\Gamma^\alpha_{\mu\nu}&=&\big\{^\alpha_{\mu\nu}\big\}+\frac{1}{2}g^{\alpha\beta}(
h_{\nu\beta}C_\mu+  h_{\mu\beta}C_\nu-
h_{\mu\nu}C_\beta),\\\nonumber
^{(2)}\Gamma^\alpha_{\mu\nu}&=&\big\{^\alpha_{\mu\nu}\big\}-\frac{1}{2}g^{\alpha\beta}(
h_{\nu\beta}C_\mu+  h_{\mu\beta}C_\nu- h_{\mu\nu}C_\beta)
\end{eqnarray}
where $h_{\mu\nu}$ is defined as the perturbed part of the metric
$g_{\mu\nu}=\bar g_{\mu\nu}+h_{\mu\nu}$ and
$\big\{^\alpha_{\mu\nu}\big\}$ is Christoffel symbol according to
$g_{\mu\nu}$. By assuming (\ref{2-connection-WG}) and considering
(\ref{generalized-connection-WG}) we have $^{(1)}\nabla_\mu
g_{\alpha\beta}=-C_\mu h_{\alpha\beta}$ and $^{(2)}\nabla_\mu
g_{\alpha\beta}=+C_\mu h_{\alpha\beta}$ where $^{(1)}\nabla_\mu$ and
$^{(2)}\nabla_\mu$ are covariant derivative with respect to
$^{(1)}\Gamma^\alpha_{\mu\nu}$ and $^{(2)}\Gamma^\alpha_{\mu\nu}$
respectively. For the above connections the average connection and
the difference tensor will be as follow respectively
\begin{eqnarray}\label{connection-tensor-WG}
\gamma^\alpha_{\mu\nu}&=&\frac{1}{2}\left(^{(1)}\Gamma^\alpha_{\mu\nu}+^{(2)}\Gamma^\alpha_{\mu\nu}\right)=\big\{^\alpha_{\mu\nu}\big\},\\\nonumber
\Omega^\alpha_{\mu\nu}&=&^{(1)}\Gamma^\alpha_{\mu\nu}-^{(2)}\Gamma^\alpha_{\mu\nu}=
g^{\alpha\beta}( h_{\nu\beta}C_\mu+  h_{\mu\beta}C_\nu-
h_{\mu\nu}C_\beta).
\end{eqnarray}
By plugging the above relations into the Lagrangian
(\ref{lag-two-conn-tensor}) or equivalently into
(\ref{lag-metric-tensor}) one gets
\begin{eqnarray}\label{lag-WG-massive}
{\cal{L}}=h_{\alpha\beta}{\cal{E}}^{\alpha\beta\mu\nu}h_{\mu\nu}-\frac{m^2}{4}\left(h_{\mu\nu}h^{\mu\nu}-h^2\right)
\end{eqnarray}
where ${\cal{E}}^{\alpha\beta\mu\nu}$ is the EGR kinetic operator,
$h=h^\mu_\mu$, $C^2=-m^2$ and $C^\mu h_{\mu\nu}=0$. The above
Lagrangian is Fierz-Pauli Lagrangian which is ghost free Lagrangian
for massive gravity at linear order (in equations of motion).
Mathematically it is ghost free because of relative coefficient
between $h_{\mu\nu}h^{\mu\nu}$ and $h^2$ is minus one. This case
happens in multi-connection framework automatically and it was not
trivial from starting point of this model. In addition we should say
in the multi-connection framework this mass term has a geometrical
meaning too. In a given distribution of connections the variance (or
standard deviation) is represented by existence of mass term. In
other words the existence of the mass term shows how far the connections are
from the average connection (Christoffel symbol).

\subsection{Non-Linear Massive Gravity}
We could show that the linear (at the level of equation of motion) Fierz-Pauli term can be deduced automatically in our setup. However it is crucial to show how one can extend the above model for non-linear dRGT massive gravity. It can be done by modifying the
connections in (\ref{2-connection-WG}) to
\begin{eqnarray}\label{2-connection-WG-higher-order}
^{(1)}\Gamma^\alpha_{\mu\nu}&=&\big\{^\alpha_{\mu\nu}\big\}+\frac{1}{2}g^{\alpha\beta}(
{\cal{K}}_{\nu\beta}C_\mu+  {\cal{K}}_{\mu\beta}C_\nu-
{\cal{K}}_{\mu\nu}C_\beta),\\\nonumber
^{(2)}\Gamma^\alpha_{\mu\nu}&=&\big\{^\alpha_{\mu\nu}\big\}-\frac{1}{2}g^{\alpha\beta}(
{\cal{K}}_{\nu\beta}C_\mu+  {\cal{K}}_{\mu\beta}C_\nu-
{\cal{K}}_{\mu\nu}C_\beta)
\end{eqnarray}
where ${\cal{K}}_{\mu\nu}=g_{\mu\nu}-\sqrt{g_{\mu\nu}-h_{\mu\nu}}$,
which reduces to ${\cal{K}}_{\mu\nu}=h_{\mu\nu}$ in the linear order
and will produce exactly a branch of the dRGT model for higher order terms \cite{massivegravity}. It
shows this framework allows to have ghost free massive gravity even
at non-linear level. Let us recall that the above forms of connections are respectively equivalent to
 $^{(1)}\nabla_\mu
g_{\alpha\beta}=-C_\mu {\cal K}_{\alpha\beta}$ and $^{(2)}\nabla_\mu
g_{\alpha\beta}=+C_\mu {\cal K}_{\alpha\beta}$ and the corresponding Lagrangian will be as following
\begin{eqnarray}\label{lag-massive-quadratic}
{\cal{L}}_{MG}=-\frac{m^2}{4}\left({\cal K}_{\mu\nu}{\cal K}^{\mu\nu}-{\cal K}^2\right).
\end{eqnarray}

Now the natural question is if one can get the higher order dRGT terms i.e. ${\cal O}({\cal K}^3)$ and ${\cal O}({\cal K}^4)$ terms. To do this we need to generalize $^{(1)}\nabla_\mu
g_{\alpha\beta}=-C_\mu {\cal K}_{\alpha\beta}$ and $^{(2)}\nabla_\mu
g_{\alpha\beta}=+C_\mu {\cal K}_{\alpha\beta}$ to higher order terms in ${\cal K}$'s. Let us assume 
\begin{eqnarray}\label{metric-compatibility-generalization}
^{(1)}\nabla_\mu
g_{\alpha\beta}=-C_\mu X_{\alpha\beta}\\\nonumber
^{(2)}\nabla_\mu
g_{\alpha\beta}=+C_\mu X_{\alpha\beta}
\end{eqnarray}
which we know, from previous section, result in a Lagrangian as ${\cal{L}}_{MG}=-\frac{m^2}{4}\left(X_{\mu\nu}X^{\mu\nu}-X^2\right)$ for $C^\mu X_{\mu\nu}=0$ and $C^2=-m^2$. Now by assuming the following definition for $X_{\mu\nu}$
\begin{eqnarray}\label{X-general}
X_{\mu\nu}&=&{\cal K}_{\mu\nu}\\\nonumber&+&\alpha\bigg[-{\cal K} {\cal K}_{\mu\nu}+2{\cal K}_{\mu\rho}{\cal K}^\rho_\nu\bigg]\\\nonumber&+&\bigg[-\frac{1}{2}\left(\beta+\alpha^2	\right)  {\cal K}^2{\cal K}_{\mu\nu}+\beta {\cal K} {\cal K}_{\mu\rho}{\cal K}^\rho_\nu-\left(3\beta+2\alpha^2\right){\cal K}_{\mu\rho}{\cal K}^\rho_\sigma{\cal K}^\sigma_\nu+\frac{1}{2}\left(3\beta+4\alpha^2\right){\cal K}_{\rho\sigma}{\cal K}^{\rho\sigma}{\cal K}_{\mu\nu} \bigg]
\end{eqnarray}
the Lagrangian ${\cal{L}}_{MG}=-\frac{m^2}{4}\left(X_{\mu\nu}X^{\mu\nu}-X^2\right)$ will be
\begin{eqnarray}\label{full-dRGT-term}
{\cal{L}}_{MG}^{dRGT}=&-&\frac{m^2}{4}\bigg({\cal K}_{\mu\nu}{\cal K}^{\mu\nu}-{\cal K}^2\bigg)\\\nonumber&-&\frac{m^2}{2}\alpha
\bigg({\cal K}^3-3{\cal K}{\cal K}_{\mu\nu}{\cal K}^{\mu\nu}+2{\cal K}_{\mu\nu}{\cal K}^\mu_\rho{\cal K}^{\rho\nu}\bigg)\\\nonumber&-&\frac{m^2}{4}\beta\bigg({\cal K}^4-6{\cal K}^2{\cal K}_{\mu\nu}{\cal K}^{\mu\nu}+8{\cal K}{\cal K}_{\mu\nu}{\cal K}^\mu_\rho{\cal K}^{\rho\nu}+3({\cal K}_{\mu\nu}{\cal K}^{\mu\nu})^2-6{\cal K}_{\mu\nu}{\cal K}^\mu_\rho{\cal K}^\rho_\sigma{\cal K}^{\sigma\nu}\bigg) + {\cal O}({\cal K}^5)
\end{eqnarray}
which has exactly the same structure of dRGT terms \cite{massivegravity,kurt}. Note that $\alpha$ and $\beta$ are two arbitrary parameters of the model.

Although the calculations are algebraic and straightforward but they are not trivial. It is because in above definition of $X_{\mu\nu}$ before fixing the coefficients of each term, for first, second and third order terms we have one, two and four possible terms respectively. But in the Lagrangian for quadratic, cubic and quartic terms we have two, three and five possible terms respectively. It means the number of variables are less than the number of equations which should be satisfied. But it is an astonishment that a solution exists for this over-determined system of equations (see Appendix for details). Hence it can be concluded that Weyl geometrical inspired model introduced in this section in multi-connection framework is consistent with all the dRGT potential terms surprisingly.

\section{Conclusions, Discussions and Open Problems}
The multi-connection framework has been introduced and then its
kinematics and dynamics have been considered. At the level of
kinematics it has been shown that the average connection plays the
same role as the connection of usual manifold with a single
connection. At the level of dynamics both the average connection and
the differences between connection show themselves. The
Einstein-Hilbert action imposes that the average connection be the
Christoffel symbol. The appearance of the difference tensor needs a
generalization of Einstein-Hilbert action. To do this many options
exist and what we assumed is the dependence of the given
connections. For a specific case we have shown that how Weyl
geometry can be used in this framework which not only makes
connections dependent but also automatically results in Fierz-Pauli
massive gravity at the quadratic level. This can be surprisingly
generalized to de Rham-Gabadadze-Tolley massive gravity. This fact is straightforward but not trivial because the system of equations is over-determined. In this
specific example the metric compatibility relation is not satisfied
by the given connections at the perturbed level. However the average
connection is the Christoffel symbol (compatible with the general
framework) and consequently it is metric compatible.

It is worth to mention that the multi-connection framework can solve
the problems in multi-metric models by construction. In
multi-connection framework dynamics, an auxiliary field exist that
can be interpreted as the metric. This interpretation is consistent
with comparison with Einstein general relativity. Naturally this
metric is the physical metric and responsible for coupling to
matter. On the other hand it has been shown that the average of
given connections is the Christoffel symbol according to this
metric. It is important because the average connection appears in
geodesic equation. Therefore everything is consistent with the equivalence
principle. So in the multi-connection framework there is just one
metric which is responsible for coupling to matter and also the
measuring distance as well as the causal structure.

\subsection{Future Perspectives}
It seems that a multi-connection framework is well-defined and does not
suffer from ambiguities in physical interpretations. One
more deep interpretation of this framework may come from comparing
this framework with Feynman path integral. In path integral all the
paths between events A and B are allowed by a weight at the level of
quantum mechanics. Then there is a path which is special and it is
the classical path. By an analogy to this in multi-connection 
scenario one can think as follow: for parallel transportation from
point A to B on a manifold all the connections are allowed in
principle. Then the average of these connections is the Christoffel
symbol that appears in the Einstein general relativity. In other
words, the Einstein general relativity is the average geometry of
multi-connection scenario. We specifically showed all the possible
connections can be assumed as representations of an imperfection in
metric compatibility relation via Weyl geometry. The challenging question
in this viewpoint is that what does play the role of weight\footnote{The
weight that appears in path integral approach.} in this case? It should be 
emphasized that this similarity is proposed as a potential clue to understand
multi-connection model more deeply. Obviously the energy regimes of Feynman path
integral quantization and multi-connection framework are ultra-violet and infra-red
scales respectively.

The other interesting topic to consider is the notion 
of geometrical curvature in the presence of more than one connection \cite{nicola}. This is out of the scope of this work but it can shed lights on our understanding of geometry of space-time.

%\newpage

\begin{acknowledgments}
We would like to thank N. Afshordi, B. Bassett, C. de Rham, N. Doroud, P.
Ferreira, G. Gibbons, S. Jalalzadeh, P. Khosravi, K. Koyama, M.
Kunz, N. Rahmanpour, S. Speziale, G. Tasinato, R. Tavakol, A.
Tolley and M. von Strauss for fruitful discussions. We are specially grateful to T. Koivisto and T. Zlosnik
for his very useful comments on the draft and also N. Sivanandam for his comments as well
as carefull reading of the draft. 
\end{acknowledgments}

\appendix
\section{Details on ${\cal O}({\cal K}^3)$ and ${\cal O}({\cal K}^4)$ terms of dRGT model}
 In this appendix we will show in details the procedure which is used to get all dRGT massive gravity terms.
In dRGT massive gravity the coefficients of the mass term are tuned such that the model becomes ghost free. Let us recall (\ref{full-dRGT-term}) i.e. the full dRGT mass term 
\begin{eqnarray}\label{full-dRGT-term-1}
{\cal{L}}_{MG}^{dRGT}=&-&\frac{m^2}{4}\bigg({\cal K}_{\mu\nu}{\cal K}^{\mu\nu}-{\cal K}^2\bigg)\\\nonumber&-&\frac{m^2}{2}\alpha
\bigg({\cal K}^3-3{\cal K}{\cal K}_{\mu\nu}{\cal K}^{\mu\nu}+2{\cal K}_{\mu\nu}{\cal K}^\mu_\rho{\cal K}^{\rho\nu}\bigg)\\\nonumber&-&\frac{m^2}{4}\beta\bigg({\cal K}^4-6{\cal K}^2{\cal K}_{\mu\nu}{\cal K}^{\mu\nu}+8{\cal K}{\cal K}_{\mu\nu}{\cal K}^\mu_\rho{\cal K}^{\rho\nu}+3({\cal K}_{\mu\nu}{\cal K}^{\mu\nu})^2-6{\cal K}_{\mu\nu}{\cal K}^\mu_\rho{\cal K}^\rho_\sigma{\cal K}^{\sigma\nu}\bigg) + {\cal O}({\cal K}^5)
\end{eqnarray}
where $\alpha$ and $\beta$ are arbitrary coefficients but as it was mentioned the other coefficients are fixed to make the model ghost free. In Weyl geometrical inspired model in multi-connection framework we have 
\begin{eqnarray}\label{metric-compatibility-generalization-1}\nonumber
^{(1)}\nabla_\mu
g_{\alpha\beta}=-C_\mu X_{\alpha\beta}\\\nonumber
^{(2)}\nabla_\mu
g_{\alpha\beta}=+C_\mu X_{\alpha\beta}
\end{eqnarray}
 which results in the potential term as ${\cal{L}}_{MG}=-\frac{m^2}{4}\left(X_{\mu\nu}X^{\mu\nu}-X^2\right)$. However in our example to make a relationship between our model and dRGT massive gravity we can define $X_{\mu\nu}$ as a function of ${\cal K}_{\alpha\beta}$ i.e. $X_{\mu\nu}({\cal K}_{\alpha\beta})$. Now let us find $X_{\mu\nu}({\cal K}_{\alpha\beta})$ order by order to get dRGT terms (\ref{full-dRGT-term-1}).
 \itemize
 \item {second order term:}
 
 This one seems obvious since by plugging $X_{\mu\nu}={\cal K}_{\mu\nu}$ into ${\cal{L}}_{MG}=-\frac{m^2}{4}\left(X_{\mu\nu}X^{\mu\nu}-X^2\right)$  what we get is exactly the second order dRGT term i.e. the first line in (\ref{full-dRGT-term-1}). However we should mention that at this order we have just one choice for $X_{\mu\nu}$ but we get both terms at the level of the Lagrangian correctly. This means it is not a trivial result though seems very easy. This fact will be clearer in the following when we consider higher order terms.
 
 \item{third order term:}
 
 To get the third order terms in (\ref{full-dRGT-term-1}) we need to assume up to second order term for $X_{\mu\nu}$ as 
 \begin{eqnarray}
X_{\mu\nu}&=&{\cal K}_{\mu\nu}+a{\cal K} {\cal K}_{\mu\nu}+b{\cal K}_{\mu\rho}{\cal K}^\rho_\nu
\end{eqnarray}
 where $a$ and $b$ are two freedom which we need to fix. If we plug the above $X_{\mu\nu}$ into ${\cal{L}}_{MG}$ we get
 \begin{eqnarray}
 X_{\mu\nu}X^{\mu\nu}-X^2=\bigg({\cal K}_{\mu\nu}{\cal K}^{\mu\nu}-{\cal K}^2\bigg)-2a{\cal K}^3+(2a-2b){\cal K}{\cal K}_{\mu\nu}{\cal K}^{\mu\nu}+2b{\cal K}_{\mu\nu}{\cal K}^\mu_\rho{\cal K}^{\rho\nu}+{\cal O}({\cal K}^4).
 \end{eqnarray}
 Now by comparing the above result and the second line in (\ref{full-dRGT-term-1}) we get three equations for two variables (i.e. it is an over-determined system of equations)
 \begin{eqnarray}
 2b=4\alpha,\hspace{1cm}2a-2b=-6\alpha,\hspace{1cm}-2a=2\alpha.
 \end{eqnarray}
 Though in principle it is an over-determined system of equations but interestingly we can solve all three equations by just having two variables $a$ and $b$. Then we have
  \begin{eqnarray}
X_{\mu\nu}&=&{\cal K}_{\mu\nu}+\alpha\bigg(-{\cal K} {\cal K}_{\mu\nu}+2{\cal K}_{\mu\rho}{\cal K}^\rho_\nu\bigg).
\end{eqnarray}
 
 \item{fourth order term:}
 
 Exactly the procedure is same as before, by taking care of the previous results we need to assume
\begin{eqnarray}
X_{\mu\nu}&=&{\cal K}_{\mu\nu}+\alpha\bigg[-{\cal K} {\cal K}_{\mu\nu}+2{\cal K}_{\mu\rho}{\cal K}^\rho_\nu\bigg]+a  {\cal K}^2{\cal K}_{\mu\nu}+b {\cal K} {\cal K}_{\mu\rho}{\cal K}^\rho_\nu+c{\cal K}_{\mu\rho}{\cal K}^\rho_\sigma{\cal K}^\sigma_\nu+d{\cal K}_{\rho\sigma}{\cal K}^{\rho\sigma}{\cal K}_{\mu\nu} 
\end{eqnarray}
where $a$, $b$, $c$ and $d$ are four variables which should be fixed. Again by plugging the above relation for $X_{\mu\nu}$ into ${\cal{L}}_{MG}$ we get
\begin{eqnarray}
 X_{\mu\nu}X^{\mu\nu}-X^2&=&\bigg({\cal K}_{\mu\nu}{\cal K}^{\mu\nu}-{\cal K}^2\bigg)+2\alpha\bigg({\cal K}^3-3{\cal K}{\cal K}_{\mu\nu}{\cal K}^{\mu\nu}+2{\cal K}_{\mu\nu}{\cal K}^\mu_\rho{\cal K}^{\rho\nu}\bigg)\\\nonumber&+&(-\alpha^2-2a){\cal K}^4+(5\alpha^2+2a-2b-2d){\cal K}^2{\cal K}_{\mu\nu}{\cal K}^{\mu\nu}+(-4\alpha^2+2b-2c){\cal K}{\cal K}_{\mu\nu}{\cal K}^\mu_\rho{\cal K}^{\rho\nu}\\\nonumber&+&(-4\alpha^2+2d)({\cal K}_{\mu\nu}{\cal K}^{\mu\nu})^2+(4\alpha^2+2c){\cal K}_{\mu\nu}{\cal K}^\mu_\rho{\cal K}^\rho_\sigma{\cal K}^{\sigma\nu} +{\cal O}({\cal K}^5).
 \end{eqnarray}
Now by comparing the above result with the third line in (\ref{full-dRGT-term-1}) we get five equations for four variables 
 \begin{eqnarray}
 -\alpha^2-2a=\beta,\hspace{.3cm}5\alpha^2+2a-2b-2d=-6\beta,\hspace{.3cm}-4\alpha^2+2d=3\beta,\hspace{.3cm}-4\alpha^2+2b-2c=8\beta\hspace{.3cm}4\alpha^2+2c=-6\beta
 \end{eqnarray}
which is again an over-determined system of equations. However interestingly one can solve all the above equations by assuming
\begin{eqnarray}
a=-\frac{1}{2}\beta-\frac{1}{2}\alpha^2,\hspace{.5cm} b=\beta, \hspace{.5cm}c=-3\beta-2 \alpha^2, \hspace{.5cm}d=\frac{3}{2}\beta+2\alpha^2.
\end{eqnarray}
So by assuming
\begin{eqnarray}\label{X-general-1}
X_{\mu\nu}&=&{\cal K}_{\mu\nu}\\\nonumber&+&\alpha\bigg[-{\cal K} {\cal K}_{\mu\nu}+2{\cal K}_{\mu\rho}{\cal K}^\rho_\nu\bigg]\\\nonumber&+&\bigg[-\frac{1}{2}\left(\beta+\alpha^2	\right)  {\cal K}^2{\cal K}_{\mu\nu}+\beta {\cal K} {\cal K}_{\mu\rho}{\cal K}^\rho_\nu-\left(3\beta+2\alpha^2\right){\cal K}_{\mu\rho}{\cal K}^\rho_\sigma{\cal K}^\sigma_\nu+\frac{1}{2}\left(3\beta+4\alpha^2\right){\cal K}_{\rho\sigma}{\cal K}^{\rho\sigma}{\cal K}_{\mu\nu} \bigg]
\end{eqnarray}
we can get all the dRGT terms correctly. We mention again that it is not a trivial result since the system of equations are over-determined.

%===============================================================================
%=============================================================
%\newpage

%=============================================================

\end{document}